\documentclass[10pt,twocolumn,journal]{IEEEtran}
\usepackage{times}
\usepackage{amsbsy}
\usepackage{amsmath}
\usepackage{latexsym}
\usepackage{amssymb}
\usepackage{times}
\usepackage{graphics}
\usepackage[ansinew]{inputenc}
\usepackage[dvips ]{graphicx}
\usepackage[dvips]{pstcol}
\input epsf
\usepackage{epic,eepic}
\usepackage{url}

\title{Reduced-Rank Space-Time Interference Suppression with Joint
Iterative Least Squares Algorithms for Spread Spectrum Systems }

\author{Rodrigo C. de Lamare and Raimundo Sampaio-Neto \\
\thanks{This work was partially funded by the
Ministry of Defence (MoD), UK, Project MoD, Contract No.
RT/COM/S/021 and CNPQ. Copyright (c) 2009 IEEE. Personal use of
this material is permitted. However, permission to use this
material for any other purposes must be obtained from the IEEE by
sending a request to pubs-permissions@ieee.org. Dr. R. C. de
Lamare is with the Communications Research Group, Department of
Electronics, University of York, York Y010 5DD, United Kingdom and
Prof. R. Sampaio-Neto is with CETUC/PUC-RIO, 22453-900, Rio de
Janeiro, Brazil. E-mails: rcdl500@ohm.york.ac.uk and
raimundo@cetuc.puc-rio.br}}

\begin{document}

\maketitle

\begin{abstract}
This paper presents novel adaptive space-time reduced-rank
interference suppression least squares algorithms based on joint
iterative optimization of parameter vectors. The proposed
space-time reduced-rank scheme consists of a joint iterative
optimization of a projection matrix that performs dimensionality
reduction and an adaptive reduced-rank parameter vector that
yields the symbol estimates. The proposed techniques do not
require singular value decomposition (SVD) and automatically find
the best set of basis for reduced-rank processing. We present
least squares (LS) expressions for the design of the projection
matrix and the reduced-rank parameter vector and we conduct an
analysis of the convergence properties of the LS algorithms. We
then develop recursive least squares (RLS) adaptive algorithms for
their computationally efficient estimation and an algorithm for
automatically adjusting the rank of the proposed scheme. A
convexity analysis of the LS algorithms is carried out along with
the development of a proof of convergence for the proposed
algorithms. Simulations for a space-time interference suppression
application with a DS-CDMA system show that the proposed scheme
outperforms in convergence and tracking the
state-of-the-art reduced-rank schemes at a comparable complexity.\\

\end{abstract}


\section{Introduction}

\PARstart{S}{pace-time} adaptive processing (STAP) techniques have
become a fundamental enabling technology of modern systems
encountered in communications \cite{paulraj}, radar and sonar
\cite{guerci,klemm}, and navigation \cite{xiong}. The basic idea
is to gather data samples from an antenna array and process them
both spatially and temporally via a linear combination of adaptive
weights. In particular, STAP algorithms have found numerous
applications in modern wireless communications based on spread
spectrum systems and code-division multiple access (CDMA)
\cite{proakis,honig&poor}. These systems implemented with direct
sequence (DS) signalling are found in third-generation cellular
telephony \cite{holma,miller,delamaretc}, indoor wireless networks
\cite{wifi}, satellite communications, ultra-wideband technology
\cite{win} and are being considered for future systems with
multi-carrier versions such as MC-CDMA and MC-DS-CDMA
\cite{juntti}, and in conjunction with multiple antennas
\cite{paulraj-mimo}. The advantages of spread spectrum systems
include good performance in multi-path channels, flexibility in
the allocation of channels, increased capacity in bursty and
fading environments and the ability to share bandwidth with
narrowband communication systems without performance degradation
\cite{proakis}.

There are numerous algorithms with different trade-offs between
performance and complexity for designing STAP techniques
\cite{haykin}. Among them, least squares (LS)-based algorithms are
often the preferred choice with respect to convergence
performance. However, when the number of filter elements in the
STAP algorithm is large they require a large number of samples to
reach its steady-state behavior and may encounter problems in
tracking the desired signal. Reduced-rank STAP techniques
\cite{scharfo}-\cite{delamarejidf} are powerful and effective
approaches in low-sample support situations and in problems with
large filters. These algorithms can effectively exploit the
low-rank nature of signals that are found in spread spectrum
communications. Their advantages are faster convergence speed and
better tracking performance than full-rank techniques when dealing
with a large number of weights. It is well known that the optimal
reduced-rank approach is based on the singular value decomposition
(SVD) of the known input data covariance matrix ${\bf R}$
\cite{scharfo}. However, this covariance matrix must be estimated.
The approach taken to estimate ${\bf R}$ and perform
dimensionality reduction is of central importance and plays a key
role in the performance of the system. Numerous reduced-rank
strategies have been proposed in the last two decades. Among the
first methods are those based on the SVD of time-averaged
estimates of ${\bf R}$ \cite{scharfo}-\cite{song&roy}, in which
the dimensionality reduction is carried out by a projection matrix
formed by appropriately selected eigenvectors computed with the
SVD. {  An effective approach to address the problem of selection
of eigenvectors, known as the cross-spectral method, and that
results in improved performance was considered in
\cite{cross-spectral}. Iterative algorithms that avoid the SVD but
do not fully exploit the structure of the data for reduced-rank
processing were reported in \cite{hua1,hua2}.} A more recent and
elegant approach to the problem was taken with the advent of the
multistage Wiener filter (MSWF) \cite{gold&reed}, which was later
extended to adaptive versions by Honig and Goldstein in
\cite{goldstein}, STAP applications \cite{hu} and other related
techniques \cite{wu}. Another method that was reported about the
same time as the MSWF is the auxiliary vector filtering (AVF)
algorithm \cite{avf}-\cite{avf5}. A reduced-rank method based on
interpolated filters with time-varying interpolators was reported
in \cite{delamaresp,delamarecl,delamaretvt} for temporal
processing and an associated STAP version was considered in
\cite{delamareiet}, however, this approach shows significant
performance degradation with small ranks. A key limitation with
the existing reduced-rank STAP techniques is the lack or a
deficiency with the exchange of information between the projection
matrix that carries out dimensionality reduction and the
subsequent reduced-rank filtering.

In this work we propose reduced-rank STAP LS algorithms for
interference suppression in spread spectrum systems. The proposed
algorithms do not require SVD and prior knowledge of the reduced
model order. The proposed reduced-rank STAP scheme consists of a
joint iterative optimization of a projection matrix that performs
dimensionality reduction and is followed by an adaptive
reduced-rank filter. The key aspect of the proposed approach is to
exchange information between the tasks of dimensionality reduction
and reduced-rank processing. {  The proposed STAP scheme builds on
the temporal scheme first reported in \cite{delamarespl07} with
stochastic gradient algorithms and extends it to the case of
spatio-temporal processing and to a deterministic
exponentially-weighted least squares design criterion}. We develop
least squares (LS) optimization algorithms and expressions for the
joint design of the projection matrix and the reduced-rank filter.
We derive recursive LS (RLS) adaptive algorithms for their
computationally efficient implementation along with a complexity
study of the proposed and existing algorithms. We also devise an
algorithm for automatically adjusting the rank of the filters
utilized in the proposed STAP scheme. A convexity analysis of the
proposed LS optimization of the filters is conducted, and an
analysis of the convergence of the proposed RLS algorithms is also
carried out. The performance of the proposed scheme is assessed
via simulations for a space-time interference suppression
application in DS-CDMA systems. The main contributions of this
work are summarized as follows: 1) A reduced-rank STAP scheme for
spatio-temporal processing of signals; 2) LS expressions and
recursive algorithms for STAP parameter estimation; 3) An
algorithm for automatically adjusting the rank of the filters; 4)
Convexity analysis and convergence proof of the proposed LS-based
algorithms.

This work is organized as follows. Section II presents the
space-time system model, and Section III states the reduced-rank
estimation problem. Section IV presents the novel reduced-rank
scheme, the joint iterative optimization and the LS design of the
filters. Section V derives the RLS and the rank adaptation
algorithms for implementing the proposed scheme. Section VI
develops the analysis of the proposed algorithms. Section VII
shows and discusses the simulations, while Section VIII gives the
conclusions.

\section{Space-Time System Model}

We consider the uplink of DS-CDMA system with symbol interval $T$,
chip period $T_c$, spreading gain $N=T/Tc$, $K$ users, { multipath
channels with $L$ propagation paths and $L < N$}. The system is
equipped with an antenna that consists of a uniform linear array
(ULA) and $J$ sensor elements \cite{guerci,klemm}. { In the model
adopted, the intersymbol interference (ISI) span and contribution
are functions of the processing gain $N$ and $L$
\cite{honig&poor}.} For instance, we assume that $L \leq N$ which
results in the interference between $3$ symbols in total, the
current one, the previous and the successive symbols. The spacing
between the ULA elements is $d=\lambda_c/2$, where $\lambda_c$ is
carrier wavelength. {  We assume that the channel is constant
during each symbol, the base station receiver is perfectly
synchronized and the delays of the propagation paths are multiples
of the chip rate.} The received signal after filtering by a
chip-pulse matched filter and sampled at the chip period yields
the $JM\times 1$ received vector at time $i$
\begin{equation}
\begin{split}
{\boldsymbol r}[i] & =  \sum_{k=1}^{K} A_{k}b_{k}[i-1]
\bar{\boldsymbol p}_{k}[i-1] + A_{k}b_{k}[i] {\boldsymbol
p}_{k}[i] \\ & \quad + A_{k}b_{k}[i+1] \tilde{\boldsymbol
p}_{k}[i+1] + {\boldsymbol
 n}[i],
\end{split}
\end{equation}
where $M=N+L-1$, the complex Gaussian noise vector is
${\boldsymbol n}[i] = [n_{1}[i] ~\ldots~n_{JM}[i]]^{T}$ with
$E[{\boldsymbol n}[i]{\boldsymbol n}^{H}[i]] =
\sigma^{2}{\boldsymbol I}$, $(\cdot)^{T}$ and $(\cdot)^{H}$ denote
transpose and Hermitian transpose, respectively, and $E[\cdot]$
stands for expected value. The spatial signatures for previous,
current and future data symbols are
\begin{equation}
\begin{split}
\bar{\boldsymbol p}_{k}[i-1] & = \boldsymbol{\bar{\mathcal F}}_{k}
\boldsymbol{{\mathcal H}}_{k}[i-1], \\ {\boldsymbol p}_{k}[i] & =
\boldsymbol{{\mathcal F}}_{k} \boldsymbol{{\mathcal H}}_{k}[i], \\
\tilde{\boldsymbol p}_{k}[i+1] & = \boldsymbol{\tilde{\mathcal
F}}_{k} \boldsymbol{{\mathcal H}}_{k}[i+1], \label{sigs}
\end{split}
\end{equation}
{  where $\boldsymbol{\bar{\mathcal F}}_{k}$,
$\boldsymbol{{\mathcal F}}_{k}$ and $\boldsymbol{\tilde{\mathcal
F}}_{k}$ are block diagonal matrices with versions of segments of
the signature sequence ${\boldsymbol s}_{k} = [a_{k}(1) \ldots
a_{k}(N)]^{T}$ of user $k$ shifted down by one position (one-chip)
and given by $\boldsymbol{\bar{\mathcal F}}_{k}  = {\rm diag}
\Big(\boldsymbol{\bar{\mathcal C}}_{k}, \boldsymbol{\bar{\mathcal
C}}_{k}, \ldots,\boldsymbol{\bar{\mathcal C}}_{k}\Big)$,
$\boldsymbol{{\mathcal F}}_{k}  = {\rm diag}
\Big(\boldsymbol{{\mathcal C}}_{k}, \boldsymbol{{\mathcal C}}_{k},
\ldots,\boldsymbol{{\mathcal C}}_{k}\Big)$,  and
$\boldsymbol{\tilde{\mathcal F}}_{k}  = {\rm diag}
\Big(\boldsymbol{\tilde{\mathcal C}}_{k},
\boldsymbol{\tilde{\mathcal C}}_{k},
\ldots,\boldsymbol{\tilde{\mathcal C}}_{k}\Big)$. The structure of
the $M \times L$ matrices $\boldsymbol{\bar{\mathcal C}}_{k}$,
$\boldsymbol{{\mathcal C}}_{k}$ and $\boldsymbol{\tilde{\mathcal
C}}_{k}$ is detailed in \cite{delamaretc}. 
The $JL \times 1$ space-time channel vector is given by
\begin{equation}
\boldsymbol{{\mathcal H}_{k}}[i] = \big[{\boldsymbol
h}_{k,0}^{T}[i] |~{\boldsymbol h}_{k,1}^{T}[i]|  ~\ldots
~|{\boldsymbol h}_{k,J-1}^{T}[i]\big]^{T}, \label{channel}
\end{equation}
where ${\boldsymbol h}_{k,l}[i] = [h_{k,0}^{(l)}[i] \ldots
h_{k,L-1}^{(l)}[i]]^{T}$ is the $L\times 1$ channel vector of user
$k$ at antenna element $l$ with their associated directions of
arrival (DoAs) $\phi_{k,m}$. The DoAs are assumed different for
each user and path \cite{gold&reed}.}

\section{Reduced-Rank STAP for Interference Suppression and Problem Statement}

In this section, we outline the main problem of STAP design for
interference suppression in spread spectrum systems and we
consider the design of reduced-rank STAP algorithms using a least
squares approach. {  The main goal of the STAP algorithms is to
jointly perform temporal filtering with spatial filtering
(beamforming) through adaptive combination of filter
coefficients.}


Let us consider the space-time received signals of the previous
section and the data organized in $JM \times 1$ vectors
${\boldsymbol r}[i]$. In order to process this data vector, one
can design a STAP algorithm that consists of a $JM \times 1$
filter ${\boldsymbol w}[i]= [w_1^{[i]}~ w_2^{[i]} ~ \ldots ~
w_{JM}^{[i]}]^T$, which adaptively and linearly combines its
coefficients with the received data samples to yield an estimate
$x[i]= {\boldsymbol w}^H[i] {\boldsymbol r}[i]$. The design of
${\boldsymbol w}[i]$ can be performed via the minimization of the
exponentially weighted LS cost function
\begin{equation}
{\mathcal{C}}({\boldsymbol w}[i] )  = \sum_{l=1}^i \lambda^{i-l} |
d[l] - {\boldsymbol w}^{H}[i]{\boldsymbol r}[l]|^2, \label{ls1}
\end{equation}
where $d[l]$ is the desired signal and $\lambda$ stands for the
forgetting factor. Solving for ${\boldsymbol w}[i]$, we obtain
\begin{equation}
{\boldsymbol w}[i] = {\boldsymbol R}^{-1}[i]{\boldsymbol p}[i],
\end{equation}
where ${\boldsymbol R}[i] = \sum_{l=1}^i \lambda^{i-l}
{\boldsymbol r}[l]{\boldsymbol r}^{H}[l]$ is the time-averaged
correlation matrix and ${\boldsymbol p}[i]=\sum_{l=1}^i
\lambda^{i-l} d^*[l]{\boldsymbol r}[l]]$ is the cross-correlation
vector.

A problem with STAP algorithms is that the laws that govern their
convergence and tracking behavior imply that the performance is a
function of $JM$, the number of elements in the filter. Thus,
large $JM$ implies slow convergence and poor tracking performance.
A reduced-rank STAP algorithm attempts to circumvent this
limitation by exploiting the low-rank nature of spread spectrum
systems and performing dimensionality reduction. This
dimensionality reduction reduces the number of adaptive
coefficients and extracts the key features of the processed data.
It is accomplished by projecting the received vectors onto a lower
dimensional subspace. Specifically, consider a $JM \times D$
projection matrix ${\boldsymbol T}_{D}[i]$ which carries out a
dimensionality reduction on the received data as given by
\begin{equation}
\bar{\boldsymbol r}[i] = {\boldsymbol T}_D^H[i] {\boldsymbol
r}[i],
\end{equation}
where in what follows all $D$-dimensional quantities are denoted
with a "bar." The resulting projected received vector
$\bar{\boldsymbol r}[i]$ is the input to a tapped-delay line
represented by the $D \time 1$ vector $\bar{\boldsymbol w}[i]=[
\bar{w}_1^{[i]} ~\bar{w}_2^{[i]}~\ldots\bar{w}_D^{[i]}]^T$. The
reduced-rank STAP output is
\begin{equation}
x[i] = \bar{\boldsymbol w}^{H}[i]\bar{\boldsymbol r}[i].
\end{equation}
If we consider the LS design in (\ref{ls1}) with the reduced-rank
parameters we obtain
\begin{equation}
\bar{\boldsymbol w}[i] = \bar{\boldsymbol
R}^{-1}[i]\bar{\boldsymbol p}[i],
\end{equation}
where $\bar{\boldsymbol R}[i] = \sum_{l=1}^i \lambda^{i-l}
\bar{\boldsymbol r}[l]\bar{\boldsymbol r}^{H}[l]={\boldsymbol
T}_D^H[i]{\boldsymbol R}[i]{\boldsymbol T}_D[i]$ is the
reduced-rank correlation matrix, $\bar{\boldsymbol
p}[i]=\sum_{l=1}^i \lambda^{i-l} d^*[l]\bar{\boldsymbol
r}[l]={\boldsymbol T}_D^H[i]{\boldsymbol p}[i]$ is the
cross-correlation vector of the reduced-rank model. The associated
sum of error squares (SES) for a rank-$D$ STAP is expressed by
\begin{equation}
\begin{split}
{\rm SES} & = \sigma^2_d - \bar{\boldsymbol p}^H[i]
\bar{\boldsymbol R}^{-1}[i]\bar{\boldsymbol p}[i] \\ & =
\sigma^2_d - {\boldsymbol p}^H[i]{\boldsymbol T}_D[i]
({\boldsymbol T}_D^H[i]{\boldsymbol R}[i]{\boldsymbol
T}_D[i])^{-1} {\boldsymbol T}_D^H[i]{\boldsymbol p}[i],
\label{ses}
\end{split}
\end{equation}
where $\sigma^2_d=\sum_{l=1}^i \lambda^{i-l} | d(l)|^2$. The
development above shows us that the key aspect for constructing
reduced-rank STAP schemes is the design of ${\boldsymbol T}_D[i]$
since the SES in (\ref{ses}) depends on ${\boldsymbol p}[i]$,
${\boldsymbol R}[i]$ and ${\boldsymbol T}_D[i]$. The quantities
${\boldsymbol p}[i]$ and ${\boldsymbol R}[i]$ are common to both
reduced-rank and full-rank STAP designs, however, the projection
matrix ${\boldsymbol T}_D[i]$ plays a key role in the
dimensionality reduction and in the performance. {  The strategy
is to find the most appropriate trade-off between the model bias
and variance \cite{scharfo} by adjusting the rank $D$ and
exchanging information between ${\boldsymbol T}_D[i]$ and
${\boldsymbol w}[i]$. {  For instance, evaluating numerically the
SES expression in equation (\ref{ses}) one can verify the
convergence and steady state performance of reduced-rank STAP
algorithms.} Next, we present the proposed reduced-rank STAP
approach.}

\section{Proposed Reduced-Rank STAP and Least Squares Design}

In this section we detail the principles of the proposed
reduced-rank STAP scheme and present a least squares (LS) design
approach for the filters. The proposed reduced-rank STAP scheme is
depicted in Fig. 1 and is formed by a projection matrix
${\boldsymbol T}_{D}[i]$ with dimensions $JM \times D$ that is
responsible for the dimensionality reduction and a $D \times 1$
reduced-rank filter $\bar{\boldsymbol w}[i]$. The $JM \times 1$
received data vector ${\boldsymbol r}[i]$ is mapped by
${\boldsymbol T}_{D}[i]$ into a $D \times 1$ reduced-rank data
vector $\bar{\boldsymbol r}[i]$. The reduced-rank filter
$\bar{\boldsymbol w}[i]$ linearly combines $\bar{\boldsymbol
r}[i]$ in order to yield a scalar estimate $x[i]$. The key
strategy of the proposed framework lies in the joint design of the
projection matrix ${\boldsymbol T}_{D}[i]$ and the reduced-rank
filter $\bar{\boldsymbol w}[i]$ according to the LS criterion. {
The exchange of information between ${\boldsymbol T}_{D}[i]$ and
$\bar{\boldsymbol w}[i]$ is different from the MSWF
\cite{gold&reed}-\cite{hu} and the AVF techniques
\cite{avf}-\cite{avf5}. {  In particular, the expressions of the
filters obtained for the proposed reduced-rank STAP scheme allow a
more efficient introduction of the bias than  that of the MSWF and
the AVF by alternating the recursions for ${\boldsymbol T}_{D}[i]$
and $\bar{\boldsymbol w}[i]$.} In addition, the proposed STAP
scheme is based on a subspace projection designed according to a
joint and iterative minimization of the LS cost function and which
achieves better performance than the Krylov subspace of the MSWF
and the AVF.}

\begin{figure}[htb]
       \centering  
       \hspace*{-3.75em}
        \vspace*{-2.0em}
      {\includegraphics[width=11.5cm, height=4.5cm]{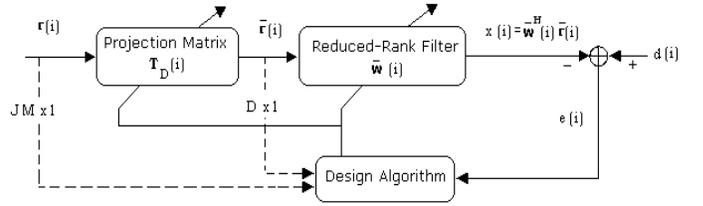}}
       \vspace*{-2em}
       \caption{\small Proposed Reduced-Rank STAP Scheme.}
\end{figure}

Let us now detail the quantities involved in the proposed
reduced-rank STAP scheme. Specifically, the projection matrix
${\boldsymbol T}_{D}[i]$ is structured as a bank of $D$ full-rank
filters with dimensions $JM \times 1$ which are described by
\begin{equation}
{\boldsymbol t}_d[i] = \big[t_{1,d}^{[i]} ~~ t_{2,d}^{[i]}~~
\ldots~~t_{JM,d}^{[i]} \big]^T, ~~ d = 1,~\ldots,~D.
\end{equation}
The filters ${\boldsymbol t}_d[i]$ are then gathered and
organized, yielding
\begin{equation}
{\boldsymbol T}_{D}[i] = \big[~{\boldsymbol t}_1^{[i]} ~|
~{\boldsymbol t}_2^{[i]}~| ~\ldots~|{\boldsymbol t}_D^{[i]}~\big],
\end{equation}
The output estimate $x[i]$ of the reduced-rank STAP scheme can be
expressed as a function of the received data ${\boldsymbol r}[i]$,
the projection matrix ${\boldsymbol T}_D[i]$ and the reduced-rank
filter $\bar{\boldsymbol w}[i]$ as given by
\begin{equation}
\begin{split}
x[i] & =  \bar{\boldsymbol w}^H[i] {\boldsymbol T}_D^H[i]
{\boldsymbol r}[i] = \bar{\boldsymbol w}^H[i] \bar{\boldsymbol
r}[i].
\end{split}
\end{equation}
Interestingly, for $D=1$, the proposed STAP scheme becomes a
conventional full-rank STAP algorithm with an additional weight
parameter $w_D$ that can be seen as a gain. For $D>1$, the signal
processing tasks are changed and the full-rank filters
${\boldsymbol t}_d[i]$ perform dimensionality reduction and the
reduced-rank filter estimates the desired signal.

In order to design the projection matrix ${\boldsymbol T}_D[i]$
and the reduced-rank filter $\bar{\boldsymbol w}[i]$ we need to
adopt an appropriate design criterion. We will resort to an
exponentially-weighted LS approach since it is mathematically
tractable and results in joint optimization algorithms that can
track time-varying signals by adjusting the forgetting factor
$\lambda$. The design of the proposed scheme amounts to solving
the following optimization problem
\begin{equation}
\begin{split}
\big[ {\boldsymbol T}_{D,{\rm opt}}[i], {\bar{\boldsymbol
w}}^{H}_{\rm opt}[i] \big] = \arg \min_{ {\boldsymbol T}_D[i],
\bar{\boldsymbol w}^{H}[i]} \sum_{l=1}^i \lambda^{i-l} | d[l] -
\bar{\boldsymbol w}^{H}[i]{\boldsymbol T}_D^H[i]{\boldsymbol
r}[l]|^2 \big], \label{prop} 
\end{split}
\end{equation}
In order to solve the above minimization problem, let us then
consider the cost function
\begin{equation}
\begin{split}
{\mathcal{C}} ( {\boldsymbol T}_D[i], \bar{\boldsymbol w}^{H}[i])
& = \sum_{l=1}^i \lambda^{i-l} | d[l] - \bar{\boldsymbol
w}^{H}[i]{\boldsymbol T}_D^H[i]{\boldsymbol
r}[l]|^2 \big], \label{propcf} 
\end{split}
\end{equation}
Minimizing (\ref{propcf}) with respect to ${\boldsymbol T}_D[i]$
we obtain
\begin{equation}
{\boldsymbol T}_{D,{\rm opt}}[i] = {\boldsymbol R}^{-1}[i]
{\boldsymbol P}_D[i] {\boldsymbol R}_{\bar{\boldsymbol
w}}^{-1}[i], \label{filtt}
\end{equation}
where ${\boldsymbol P}_D[i] = \sum_{l=1}^i
\lambda^{i-l}d^{*}[l]{\boldsymbol r}[l]\bar{\boldsymbol w}^H[i]$,
the time-averaged correlation matrix is ${\boldsymbol R}[i] =
\sum_{l=1}^i \lambda^{i-l}{\boldsymbol r}[l]{\boldsymbol
r}^{H}[l]$ and ${\boldsymbol R}_{\bar{\boldsymbol w}}[i] =
{\bar{\boldsymbol w}}[i]{\bar{\boldsymbol w}}^{H}[i]$. {  Note
that we have opted for computing ${\boldsymbol
R}_{\bar{\boldsymbol w}}[i]$ as ${\boldsymbol R}_{\bar{\boldsymbol
w}}[i] = \sum_{l=1}^i \lambda^{i-l}{\bar{\boldsymbol
w}}[l]{\bar{\boldsymbol w}}^{H}[l]$ with a regularization term
introduced at the beginning of the iterations in order to allow
the computation of its inverse. For this reason, the latter
approach will be adopted for the derivation of adaptive
algorithms.} Minimizing (\ref{propcf}) with respect to
$\bar{\boldsymbol w}[i]$, the reduced-rank filter becomes
\begin{equation}
{\bar{\boldsymbol w}}_{\rm opt}[i] = \bar{\boldsymbol R}^{-1}[i]
\bar{\boldsymbol p}[i], \label{filtw}
\end{equation}
where $\bar{\boldsymbol p}[i] = {\boldsymbol T}_{D,{\rm opt}}^H[i]
\sum_{l=1}^i \lambda^{i-l}d^{*}[l]{\boldsymbol r}[l] =
\sum_{l=1}^i \lambda^{i-l}d^{*}[l]\bar{\boldsymbol r}[l]]$,
$\bar{\boldsymbol R}[i] = {\boldsymbol T}_{D,{\rm
opt}}^H[i]\sum_{l=1}^i \lambda^{i-l}{\boldsymbol r}[l]{\boldsymbol
r}^H[l] {\boldsymbol T}_{D,{\rm opt}}[i] $. The associated SES for
the proposed reduced-rank STAP scheme is
\begin{equation}
{\rm SES} = \sigma^{2}_{d} - \bar{\boldsymbol p}^{H}[i]
\bar{\boldsymbol R}^{-1}[i] \bar{\boldsymbol p}[i], \label{ses2}
\end{equation}
where $\sigma^{2}_{d}=\sum_{l=1}^i \lambda^{i-l}|d[l]|^{2}$. Note
that the expressions in (\ref{filtt}) and (\ref{filtw}) are not
closed-form solutions for ${\bar{\boldsymbol w}}_{\rm opt}[i]$ and
${\boldsymbol T}_{D,{\rm opt}}[i]$ since (\ref{filtt}) is a
function of ${\bar{\boldsymbol w}}_{\rm opt}[i]$  and
(\ref{filtw}) depends on and ${\boldsymbol T}_{D,{\rm opt}}[i]$.
Therefore they have to be iterated with an initial guess to obtain
a solution. The expressions in (\ref{filtt}) and (\ref{filtw})
require the inversion of matrices, which entails cubic complexity
with $JM$ and $D$. {  Computing the SES in (\ref{ses2}), it can be
numerically verified the convergence and steady state performances
of reduced-rank STAP algorithms, namely, the proposed, the MSWF
\cite{goldstein} and the AVF \cite{avf5}. In order to reduce the
complexity, we will derive RLS algorithms in the next section. The
rank $D$ must be set by the designer to ensure appropriate
performance or a mechanism for automatically adjusting the rank
should be adopted. We will also present an automatic rank
selection algorithm in what follows.}

\section{Proposed RLS and Rank Selection Algorithms}

In this section we propose RLS algorithms for efficiently
implementing the LS design of the previous section and estimating
the filters ${\boldsymbol T}_{D,{\rm opt}}[i]$ and
${\bar{\boldsymbol w}}_{\rm opt}[i]$ with the filters
${\boldsymbol T}_{D}[i]$ and ${\bar{\boldsymbol w}}[i]$,
respectively. We also develop rank selection algorithms for
automatically adjusting the rank $D$ of the proposed STAP
algorithm. An analysis of the computational requirements of the
proposed and analyzed algorithms is also included.

\subsection{Proposed RLS Algorithm}

In order to derive an RLS algorithm for the proposed scheme, we
consider (\ref{filtt}) and derive a recursive procedure for
computing the parameters of ${\boldsymbol T}_D[i]$. Let us define
\begin{equation}
\begin{split}
{\boldsymbol P}[i] & ={\boldsymbol R}^{-1}[i] , \\
{\boldsymbol Q}_{\bar{\boldsymbol w}}[i] & ={\boldsymbol
R}^{-1}_{\bar{\boldsymbol w}}[i-1],\\
{\boldsymbol P}_D [i] &= \lambda {\boldsymbol P}_{D}[i-1] + d^*[i]
{\boldsymbol r}[i] {\bar{\boldsymbol w}}^H [i],
\end{split}
\end{equation}
and rewrite the expression in (\ref{filtt}) as follows
\begin{equation}
\begin{split}
{\boldsymbol T}_D [i] & =  {\boldsymbol P}[i] {\boldsymbol P}_D[i] {\boldsymbol Q}_{\bar{\boldsymbol w}}[i] \\
& = \lambda {\boldsymbol P}[i] {\boldsymbol P}_D [i-1]
{\boldsymbol Q}_{\bar{\boldsymbol w}}[i] + d^*[i] {\boldsymbol
P}[i] {\boldsymbol r}[i] {\bar{\boldsymbol w}}^H[i] {\boldsymbol
Q}_{\bar{\boldsymbol w}}[i]
\\ & = {\boldsymbol T}_D[i-1] -
{\boldsymbol k}[i] {\boldsymbol P}[i-1] {\boldsymbol P}_D[i-1] {\boldsymbol Q}_{\bar{\boldsymbol w}}[i] \\
& \quad + d^*[i]{\boldsymbol P}[i]
{\boldsymbol r}[i] {\bar{\boldsymbol w}}^H[i] {\boldsymbol Q}_{\bar{\boldsymbol w}}[i] \\
& = {\boldsymbol T}_D[i-1] - {\boldsymbol k}[i] {\boldsymbol
P}[i-1] {\boldsymbol P}_D[i-1] {\boldsymbol Q}_{\bar{\boldsymbol w}}[i] \\
& \quad + d^*[i]{\boldsymbol k}[i]
 {\bar{\boldsymbol w}}^H[i] {\boldsymbol Q}_{\bar{\boldsymbol w}}[i] \\
\end{split}
\end{equation}
By defining the vector ${\boldsymbol t}[i] = {\boldsymbol
Q}_{\bar{\boldsymbol w}}[i] {\bar{\boldsymbol w}}[i]$ and using
the fact that $\bar{\boldsymbol r}^H[i-1] = {\boldsymbol
r}^H[i-1]{\boldsymbol T}_D[i-1]$ we arrive at
\begin{equation}
{\boldsymbol T}_D[i] = {\boldsymbol T}_D[i-1] + {\boldsymbol k}[i]
\big( d^*[i] {\boldsymbol t}^H[i] - \bar{\boldsymbol r}^H[i]
\big), \label{filttrec}
\end{equation}
where the Kalman gain vector for the computation of ${\boldsymbol
T}_D [i]$ is
\begin{equation}
{\boldsymbol k}[i] = \frac{\lambda^{-1} {\boldsymbol P}[i-1]
{\boldsymbol r}[i] }{1 + \lambda^{-1} {\boldsymbol r}^H[i]
{\boldsymbol P}[i-1]{\boldsymbol r}[i]}
\end{equation}
and the update for the matrix ${\boldsymbol P}[i]$ employs the
matrix inversion lemma \cite{haykin}
\begin{equation}
{\boldsymbol P}[i] = \lambda^{-1} {\boldsymbol P}[i-1] -
\lambda^{-1} {\boldsymbol k}[i] {\boldsymbol r}^H[i] {\boldsymbol
P}[i-1]
\end{equation}
the vector ${\boldsymbol t}[i]$ is updated as follows
\begin{equation}
{\boldsymbol t}[i] = \frac{\lambda^{-1} {\boldsymbol
Q}_{\bar{\boldsymbol w}}[i-1] {\bar{\boldsymbol w}}[i-1] }{1 +
\lambda^{-1} {\bar{\boldsymbol w}}^H[i-1] {\boldsymbol
Q}_{\bar{\boldsymbol w}}[i-1] {\bar{\boldsymbol w}}[i-1]}
\end{equation}
and the matrix inversion lemma is used to update ${\boldsymbol
Q}_{\bar{\boldsymbol w}}[i]$ as described by
\begin{equation}
{\boldsymbol Q}_{\bar{\boldsymbol w}}[i] = \lambda^{-1}
{\boldsymbol Q}_{\bar{\boldsymbol w}}[i-1] - \lambda^{-1}
{\boldsymbol t}[i] {\bar{\boldsymbol w}}^H[i-1]{\boldsymbol
Q}_{\bar{\boldsymbol w}}[i-1], \label{Qrec}
\end{equation}
The equations (\ref{filttrec})-(\ref{Qrec}) constitute the first
part of the proposed RLS algorithm and are responsible for
calculating the projection matrix ${\boldsymbol T}_D[i]$.

In order to derive a recursive update equation for the
reduced-rank filter ${\bar{\boldsymbol w}}[i]$, we consider the
expression in (\ref{filtw}) with its associated quantities, i.e.,
the matrix $\bar{\boldsymbol R}[i] = \sum_{l=1}^i
\lambda^{i-l}\bar{\boldsymbol r}[l] \bar{\boldsymbol r}^{H}[l]$
and the vector $\bar{\boldsymbol p}[i] = \sum_{l=1}^i
\lambda^{i-l}d^{*}[l]\bar{\boldsymbol r}[l]$. Let us define
\begin{equation}
\begin{split}
\boldsymbol{\bar{\Phi}}[i] & = {\boldsymbol R}^{-1}[i], \\
\bar{\boldsymbol p}[i] &= \lambda \bar{\boldsymbol p}[i-1] +
d^*[i] \bar{\boldsymbol r}[i],
\end{split}
\end{equation}
and then we can rewrite (\ref{filtw}) in the following alternative
form
\begin{equation}
\begin{split}
{\bar{\boldsymbol w}}[i] & = 
{\bar{\boldsymbol w}}[i-1] + \bar{\boldsymbol k}[i] \big[
d^*[i] - \bar{\boldsymbol r}^H[i] {\bar{\boldsymbol w}}[i-1] \big]
\end{split}
\end{equation}
By defining $\xi[i] = d[i] - {\bar{\boldsymbol
w}}^H[i-1]\bar{\boldsymbol r}^H[i] $ we arrive at the proposed RLS
algorithm for obtaining ${\bar{\boldsymbol w}}[i]$
\begin{equation}
{\bar{\boldsymbol w}}[i] = {\bar{\boldsymbol w}}[i-1] +
\bar{\boldsymbol k}[i] \xi^*[i], \label{filtwrec}
\end{equation}
where the so-called Kalman gain vector is given by
\begin{equation}
\bar{\boldsymbol k}[i] = \frac{\lambda^{-1}
\boldsymbol{\bar{\Phi}}[i-1] \bar{\boldsymbol r}[i] }{1 +
\lambda^{-1} \bar{\boldsymbol r}^H[i]
\boldsymbol{\bar{\Phi}}[i-1]\bar{\boldsymbol r}[i]}
\end{equation}
and the update for the matrix inverse $\boldsymbol{\bar{\Phi}}[i]$
employs the matrix inversion lemma \cite{haykin}
\begin{equation}
\boldsymbol{\bar{\Phi}}[i] = \lambda^{-1}
\boldsymbol{\bar{\Phi}}[i-1] - \lambda^{-1} \bar{\boldsymbol k}[i]
\bar{\boldsymbol r}^H[i] \boldsymbol{\bar{\Phi}}[i-1].
\label{filtphi}
\end{equation}
It should be noted that the proposed RLS algorithm given in
(\ref{filtwrec})-(\ref{filtphi}) is similar to the conventional
RLS algorithm \cite{haykin}, except that it works in a
reduced-rank model with a $D \times 1$ input $\bar{\boldsymbol r}
[i] = {\boldsymbol T}_D^H [i] {\boldsymbol r}[i]$, where the $JM
\times D$ matrix ${\boldsymbol T}_D$ is the projection matrix
responsible for dimensionality reduction.

\subsection{Rank Selection Algorithm}

{The performance of the RLS algorithm described in the previous
subsection depends on the rank $D$. This motivates the development
of methods to automatically adjust $D$ on the basis of the cost
function. Unlike prior methods for rank selection which utilize
MSWF-based algorithms \cite{goldstein} or the cross-validation
approach used with AVF-based recursions \cite{avf5}, we focus on
an approach that determines $D$ based on the LS criterion computed
by the filters ${\boldsymbol T}_D[i]$ and $\bar{\boldsymbol
w}^{(D)}[i]$, where the superscript $^{(D)}$ denotes the rank used
for the adaptation. Although there are similarities between the
algorithm described here and the one reported in \cite{goldstein},
the algorithm presented here differs from \cite{goldstein} in that
it clearly details the strategy for updating ${\boldsymbol
T}_D[i]$ and $\bar{\boldsymbol w}^{(D)}[i]$, defines {  the
maximum ($D_{\rm max}$) and minimum ($D_{\rm min}$)} values for
the rank $D$ allowed and works with extended filters for reduced
complexity. The method for automatically selecting the rank of the
algorithm is based on the exponentially weighted \textit{a
posteriori} least-squares type cost function described by
\begin{equation}
{\mathcal C}_{\rm ap}({\boldsymbol T}_D[i], {\bar{\boldsymbol
w}}^{(D)}[i]) = \sum_{l=1}^{i} \alpha^{i-l} \big|d[l]-
{\bar{\boldsymbol w}}^{H, ~ (D)}[i]{{\boldsymbol
T}}_D[i]{\boldsymbol r}[l]|^2 , \label{eq:costadap}
\end{equation}
where $\alpha$ is the forgetting factor and ${\bar{\bf
w}}^{(D)}[i]$ is the reduced-rank filter with rank $D$. For each
time interval $i$, we can select $D$ which minimizes ${\mathcal
C}({\boldsymbol T}_D[i],\bar{\boldsymbol w}^{(D)}[i])$ and the
exponential weighting factor $\alpha$ is required as the optimal
rank varies as a function of the data record. {  The dimensions of
${\boldsymbol T}_D[i]$ and $\bar{\boldsymbol w}^{(D)}[i]$ are
extended to $M \times D_{\rm max}$ and $D_{\rm max}$,
respectively,} and the associated matrices $\hat{\bar{\boldsymbol
R}}[i]$, ${\boldsymbol P}_D[i]$ and ${\boldsymbol
Q}_{\bar{\boldsymbol w}}[i]$ should be compatible for adaptation.
Our strategy is to consider the adaptation with the maximum
allowed rank $D_{\rm max}$ and then perform a search with the aim
of finding the best rank within the range $D_{\rm min}$ to $D_{\rm
max}$. To this end, we define ${\boldsymbol T}_D[i]$ and
${\bar{\boldsymbol w}}^{(D)}[i]$ as follows:
\begin{equation}
\begin{split}
{\boldsymbol T}_D[i] & = \left[\begin{array}{cccccc} {\boldsymbol
t}_1[i] &  \ldots & {\boldsymbol t}_{{\rm I_{\rm min}}}[i] &
\ldots & {\boldsymbol
t}_{\rm I_{{\rm max}}}[i] \end{array}\right]^T \\
{\bar{\boldsymbol w}}^{(D)}[i] & = \left[\begin{array}{cccccc}
{\bar w}_1[i] &  \ldots & {\bar w}_{D_{\rm min}}[i] & \ldots &
{\bar w}_{D_{\rm max}}[i]
\end{array}\right]^T
\end{split}
\end{equation}
The proposed rank selection algorithm is given by
\begin{equation}
D_{\rm opt}[i] = \arg \min_{D_{\rm min} \leq d \leq D_{\rm max}}
{\mathcal C}_{\rm ap}({\boldsymbol T}_D[i],\bar{\boldsymbol
w}^{(D)}[i]), \label{dsel}
\end{equation}
where $d$ is an integer, $D_{\rm min}$ and $D_{\rm max}$ are the
minimum and maximum ranks allowed for the reduced-rank filter,
respectively. Note that a smaller rank may provide faster
adaptation during the initial stages of the estimation procedure
and a greater rank usually yields a better steady-state
performance. Our studies reveal that the range for which the rank
$D$ of the proposed algorithms have a positive impact on the
performance of the algorithms is limited, being from $D_{\rm
min}=3$ to $D_{\rm max}=8$ for the reduced-rank filter recursions.
These values are rather insensitive to the system load (number of
users), to the number of array elements and work very well for all
scenarios and algorithms examined. The computational complexity of
the proposed rank selection algorithm with extended filters is
equivalent to the computation of the cost function in
(\ref{eq:costadap}) and requires $3(D_{\rm max} -D_{\rm min}) +1$
additions and a sorting algorithm to find the best rank according
to (\ref{dsel}). An alternative strategy to using extended filters
is the deployment of multiple filters with the rank selection
algorithm in (\ref{dsel}) that determines the best set of filters
for each time interval. Specifically, this approach employs
$D_{\rm max} - D_{\rm min}+1$ pairs of filters and has a very high
complexity.

A second approach that can be used is a mechanism based on the
observation of the columns of ${\boldsymbol T}_D[i]$ and a
stopping rule, as reported in \cite{goldstein}. The method
performs the following optimization
\begin{equation}
D_{\rm opt}[i] = \arg \max_{d} \frac{||P_{{\boldsymbol
T}_d}({\boldsymbol t}_d[i]) ||}{||{\boldsymbol t}_d[i]||} >
\delta,
\end{equation}
where $P_{{\boldsymbol T}_d} ({\boldsymbol x} ) $ is the
orthogonal projection of the vector ${\boldsymbol x}$ onto the
subspace ${\boldsymbol T}_d$ and $\delta$ is a small positive
constant. In \cite{goldstein}, it has not been discussed the use
of a range of values for allowing the selection, however, we found
that it is beneficial in terms of complexity to restrict the
optimization to an appropriate range of values $D_{\rm max}$ to
$D_{\rm min}$ as with the previous method.

Another possibility for rank selection is the use of the
cross-validation (CV) method reported in \cite{avf5}. This
approach selects the filters' lengths which minimize a cost
function that is estimated based on observations (training data)
that have not been used in the process of building the filters
themselves as described by
\begin{equation}
{\mathcal C}_{\rm CV}({\boldsymbol T}_D[i], {\bar{\boldsymbol
w}}^{(D)}[i]) = \sum_{l=1}^{i} \alpha^{i-l} \big|d(l) -
{\bar{\boldsymbol w}}^{H, ~ (D)}_{(i/l)}[i]{{\boldsymbol
T}}_{D,(i/l)}[i]{\boldsymbol r}[l]|^2 , \label{eq:costcv}
\end{equation}
We consider here the same "leave one out" approach as in
\cite{avf5}. For a given data record of size $i$, the CV approach
chooses the filter ${\bar{\boldsymbol w}}^{H, ~ (D)}[i]$ that
performs the following optimization
\begin{equation}
 D_{\rm opt} [i] = \arg \min_{ d \in \{1, 2, \ldots \}
 } {\mathcal C}_{\rm CV}({\boldsymbol T}_d[i], \bar{\boldsymbol
w}^{(d)}[i]) ,
\end{equation}
The main difference between this and the other algorithms
presented lies in the use of CV, which leaves one sample out in
the process, and the use of the constraint on the allowed filter
lengths. In the simulations, we will compare the rank selection
algorithms and discuss their advantages and disadvantages. }

\subsection{Computational Complexity}

{  In this part of the work, we detail the computational
complexity requirements of the proposed RLS algorithms and compare
them with those of existing algorithms. We also provide the
computation complexity of the proposed and existing rank selection
algorithms. The computational complexity expressed in terms of
additions and multiplications is depicted in Table I for the RLS
algorithms, the complexity of the proposed rank selection
algorithm with multiple filters including the proposed RLS
algorithm is illustrated in Table II, and that of the remaining
rank selection techniques is given in Table III.

\begin{table}[t]
\centering%
\caption{\small Computational complexity of RLS algorithms.} {
\begin{tabular}{lcc}
\hline
{\small Algorithm} & {\small Additions} & {\small Multiplications} \\
\emph{\small \bf Full-rank }\cite{haykin} & {\small $3(JM)^2 - 2JM
+ 3$} & {\small $6(JM)^{2}+2JM + 2$}
\\ \\
\emph{\small \bf Proposed}  & {\small $ 3(JM)^2 - 2JM + 3 $}  & {\small $ 7(JM)^2 + 2JM $}  \\
\emph{\small \bf }  & {\small $6D^2 - 8D + 3$}  & {\small $ 7D^2+9D$}
\\\\
\emph{\small \bf MSWF }\cite{goldstein} & {\small $D(JM)^2 +
(JM)^2
+ 6D^2 $} & {\small $D(JM)^2+(JM)^2$} \\
\emph{\small } & {\small $-8D+2$} & {\small $2DJM+3D+2$} \\\\
\emph{\small \bf AVF }\cite{avf5} & {\small $D((JM)^2+3(JM-1)^2)$} & {\small $D(4(JM)^2+4JM + 1)$} \\
\emph{\small \bf  } & {\small $+D(5(JM-1)+1)$} & {\small $4JM +2$}
\\\emph{\small  } & {\small $2JM-1$} & {\small } \\\hline
\end{tabular}
}
\end{table}

\begin{figure}[!htb]
\begin{center}
\def\epsfsize#1#2{1\columnwidth}
\epsfbox{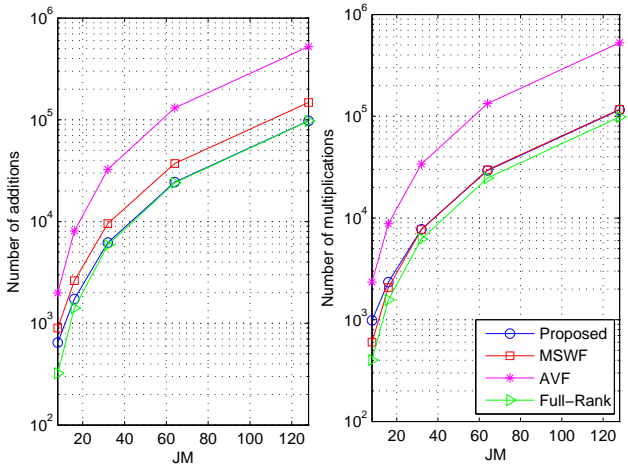} \caption{\small Complexity in terms of additions
and multiplications against number of input samples (JM) and $D=4$
.} \label{comp}
\end{center}
\end{figure}

In the case of the proposed reduced-rank RLS algorithm the
complexity is quadratic with $(JM)^2$ and $D^2$. This corresponds
to a complexity slightly higher than the one observed for the
full-rank RLS algorithm, provided $D$ is significantly smaller
than $JM$, and comparable to the cost of the MSWF-RLS
\cite{goldstein} and the AVF \cite{avf5}. In order to illustrate
the main trends in what concerns the complexity of the proposed
and analyzed algorithms, we show in Fig. \ref{comp} the complexity
against the number of input samples $JM$. The curves indicate that
the proposed reduced-rank RLS algorithm has a complexity lower
than the MSWF-RLS algorithm \cite{goldstein} and the AVF
\cite{avf5}, whereas it remains at the same level of the full-rank
RLS algorithm.

\begin{table}[t]
\centering%
\caption{\small Computational complexity of the proposed rank
selection algorithm with multiple filters.} {
\begin{tabular}{lcc}
\hline
\emph{\small \bf  } & {\small $2(D_{\rm max}- D_{\rm min}) +1$} & {\small $(D_{\rm max}-D_{\rm min}+1)\times$} \\
\emph{\small \bf  Proposed with} & {\small $(D_{\rm max}-D_{\rm min}+1)\times$} & {\small $(7(JM)^2 + 2JM$} \\
\emph{\small \bf Multiple Filters } & {\small $ (3(JM)^2 - 2JM + 3 $} & {\small $+ 7D_{\rm max}^2+9D_{\rm max})$} \\
\emph{\small \bf  } & {\small $ +6D_{\rm max}^2 - 8D_{\rm max} +
3)$} &
{\small } \\
 \hline
\end{tabular}
}
\end{table}

The proposed rank selection algorithm with multiple filters has a
number of arithmetic operations that is substantially higher than
the other compared methods since it requires the computation of
$D_{\rm max} - D_{\rm min}+1$ pairs of filters with the proposed
RLS algorithms simultaneously. We show the overall cost of this
algorithm separately in Table II. The computational complexity of
the remaining rank selection algorithms including the proposed and
the existing rank selection algorithms is shown in Table III. From
Table III, we can notice that the proposed rank selection
algorithm with extended filters is significantly less complex than
the existing methods based on projection with stopping rule
\cite{goldstein} and the CV approach \cite{avf5}. Specifically,
the proposed rank selection algorithm with extended filters only
requires $2(D_{\rm max}-D_{\rm min})$ additions, as depicted in
the first row of Table III. To this cost we must add the
operations required by the proposed RLS algorithm, whose
complexity is shown in the second row of Table I using $D_{\rm
max}$ according to the procedure outlined in the previous
subsection. The complexities of the MSWF and the AVF algorithms
are detailed in the third and fourth rows of Table I. For their
operation with rank selection algorithms, a designer must add
their complexities in Table I to the complexity of the rank
selection algorithm of interest, as shown in Table III.

\begin{table}[t]
\centering%
\caption{\small Computational complexity of remaining rank
selection algorithms.} {
\begin{tabular}{lcc}
\hline
{\small Algorithm} & {\small Additions} & {\small Multiplications} \\
\emph{\small \bf Proposed with}  & {\small $ 2(D_{\rm max}- D_{\rm min}) +1$}  & {\small $ -$}  \\
\emph{\small \bf Extended Filter }  & {\small $ $}  & {\small $ $}

\\\\
\emph{\small \bf Projection with } & {\small
$2(2JM-1)\times $} & {\small $((JM)^2+JM+1)\times $} \\
\emph{\small \bf Stopping Rule }\cite{goldstein} & {\small
$(D_{\rm max}- D_{\rm min})+1$} & {\small $(D_{\rm max}-D_{\rm
min}+1)$}
\\\\
\emph{\small \bf CV }\cite{avf5} & {\small $(2JM-1)\times $} & {\small $(D_{\rm max}- D_{\rm min}+1)\times$} \\
\emph{\small \bf  } & {\small $(2(D_{\rm max}- D_{\rm min}) +1)$}
& {\small $JM + 1$}
\\
\hline
\end{tabular}
}
\end{table}

}

\section{Analysis}

In this section, we conduct a convexity analysis of the proposed
optimization that is responsible for designing the filters
${\boldsymbol T}_{D}[i]$ and ${\bar{\boldsymbol w}}[i]$ of the
proposed scheme. We show that the proposed optimization leads to a
problem with multiple solutions, and we discuss the properties of
the method. In particular, we conjecture that it leads to a
problem with multiple and possibly identical minimum points. This
is corroborated by numerous studies that verify that the method is
insensitive to different initializations (except for the case when
${\boldsymbol T}_{D}[i]$ is a null matrix and which annihilates
the received signal) and that is always converge to the same point
of minimum. We also establish the convergence of the proposed
optimization algorithm, showing that the sequence of filters
${\boldsymbol T}_{D}[i]$ and ${\bar{\boldsymbol w}}[i]$ produces a
sequence of outputs that is bounded.

\subsection{Convexity Analysis of the Proposed Method}

In this part, we carry out a convexity analysis of the proposed
reduced-rank scheme and LS optimization algorithm. Our approach is
based on expressing the output of the proposed scheme in a
convenient form that renders itself to analysis. Let us consider
the proposed optimization method in (\ref{prop}) and express it by
an expanded cost function
\begin{equation}
\begin{split}
{\mathcal{C}} ( {\boldsymbol T}_D[i], \bar{\boldsymbol w}^{H}[i])
& = \sum_{l=1}^i \lambda^{i-l} | d[l] - \bar{\boldsymbol
w}^{H}[i]{\boldsymbol T}_D^H[i]{\boldsymbol r}[l]|^2 \big]\\
& = \sum_{l=1}^i \lambda^{i-l} | d[l]|^2  - \sum_{l=1}^i
\lambda^{i-l} \bar{\boldsymbol w}^{H}[i] {\boldsymbol
T}_D^H[i]d^*[l]{\boldsymbol r}[l] \\ & \quad - \sum_{l=1}^i
\lambda^{i-l} d[l]{\boldsymbol r}^H[l]  {\boldsymbol
T}_D[i]\bar{\boldsymbol w}[i] \\ & \quad + \sum_{l=1}^i
\lambda^{i-l} \bar{\boldsymbol w}^{H}[i]{\boldsymbol
T}_D^H[i]{\boldsymbol r}[l]{\boldsymbol r}^H[l]  {\boldsymbol
T}_D[i]\bar{\boldsymbol w}[i], \label{propcf3}
\end{split}
\end{equation}
In order to proceed, let us express $x[i]$ in an alternative and
more convenient form as
\begin{equation}
\begin{split}
x[i] &  = \bar{\boldsymbol w}^H[i] {\boldsymbol T}_D^H[i]
{\boldsymbol r}[i] =  \bar{\boldsymbol w}^H[i] \sum_{d=1}^D
{\boldsymbol T}_D^H[i] {\boldsymbol r}[i] {\boldsymbol v}_d \\ & =
\bar{\boldsymbol w}^H[i] \left[\begin{array}{cccc}
{\boldsymbol r}[i] & 0 & \ldots & 0 \\
0 & {\boldsymbol r}[i]  & \ddots & 0 \\
\vdots & \vdots  & \ddots & \vdots \\
0 & \ldots & \ldots  & {\boldsymbol r}[i]
\end{array} \right]^T \left[\begin{array}{c}
{\boldsymbol s}_1^*[i] \\ {\boldsymbol s}_2^*[i] \\ \vdots \\
{\boldsymbol s}_D^*[i]
\end{array} \right] \\ & = \bar{\boldsymbol w}^H[i] {\boldsymbol
\Re}^T[i] {\boldsymbol s}_v^*[i]
\end{split}
\end{equation}
where ${\boldsymbol \Re}[i]$ is a $DJM \times D$ block diagonal
matrix with the input data vector ${\boldsymbol r}[i]$,
${\boldsymbol s}_v^*[i]$ is a $DJM \times 1$ vector with the
columns of ${\boldsymbol T}_D[i]$ stacked on top of each other and
and the $D \times 1$ vector ${\boldsymbol v}_d$ contains a $1$ in
the $d$-th position and zeros elsewhere.

In order to analyze the proposed joint optimization procedure, we
can rearrange the terms in $x[i]$ and define a single $D(JM+1)
\times 1$ parameter vector ${\boldsymbol q}[i] = [\bar{\boldsymbol
w}^T[i] ~ {\boldsymbol s}_v^T[i]]^T$. We can therefore further
express $x[i]$ as
\begin{equation}
\begin{split}
x[i] &  = {\boldsymbol q}^H[i] \left[\begin{array}{cc}
{\boldsymbol 0}_{D \times 1} & {\boldsymbol 0}_{D \times DJM} \\
{\boldsymbol \Re}[i] & {\boldsymbol 0}_{DJM \times DJM}
\end{array} \right] {\boldsymbol q}[i] \\ & = {\boldsymbol q}^H[i]
{\boldsymbol U}[i] {\boldsymbol q}[i]
\end{split}
\end{equation}
where ${\boldsymbol U}[i]$ is a $D(JM+1) \times D(JM+1)$ matrix
which contains ${\boldsymbol \Re}[i]$. At this stage, we can
alternatively express the cost function in (\ref{propcf3}) as
\begin{equation}
\begin{split}
{\mathcal{C}} ( {\boldsymbol q}[i] ) & = \sum_{l=1}^i
|d[l]-{\boldsymbol q}^H[i]{\boldsymbol U}[l] {\boldsymbol
q}[i]|^2. \label{uopt3}
\end{split}
\end{equation}
We can examine the convexity of the above by computing the Hessian
(${\boldsymbol H}$)with respect to ${\boldsymbol q}[i]$ using the
expression \cite{luen}
\begin{equation}
{\boldsymbol H} = \frac{\partial}{\partial {\boldsymbol q}^H[i]}
\frac{\partial ({\mathcal{C}} ( {\boldsymbol q}[i] )}{\partial
{\boldsymbol q}[i]} \label{hess}
\end{equation}
and testing if the terms are positive semi-definite. Specifically,
${\boldsymbol H}$ is positive semi-definite if ${\boldsymbol
a}^{H}{\boldsymbol H}{\boldsymbol a} \geq 0$ for all nonzero
${\boldsymbol a} \in \boldsymbol{C}^{D(JM+1)\times D(JM+1)}$
\cite{luen,golub}. Therefore, the optimization problem is convex
if the Hessian ${\boldsymbol H}$ is positive semi-definite.

Evaluating the partial differentiation in the expression given in
(\ref{hess}) yields
\begin{equation}
\begin{split}
{\boldsymbol H} & = \sum_{l=1}^i ({\boldsymbol q}^H[i]
{\boldsymbol U}[l] {\boldsymbol q}[i] -d^*[l]) {\boldsymbol U}[l]
+
\sum_{l=1}^i {\boldsymbol U}^H[l] {\boldsymbol q}[i] {\boldsymbol q}[i]^H{\boldsymbol U}[l] \\
& \quad  +  \sum_{l=1}^i ({\boldsymbol q}^H[i] {\boldsymbol U}[l]
{\boldsymbol q}[i] -d[l]) {\boldsymbol U}^H[l]  +
\sum_{l=1}^i {\boldsymbol U}[l] {\boldsymbol q}[i] {\boldsymbol q}[i]^H{\boldsymbol U}^H[l] \\
\end{split}
\end{equation}
By examining ${\boldsymbol H}$, we verify that the second and
fourth terms are positive semi-definite, whereas the first and the
third terms are indefinite. Therefore, the optimization problem
can not be classified as convex. It is however important to remark
that our studies indicate that there are no local minima and there
exists multiple solutions (which are conjectured to be identical).

In order to support this claim, we have checked the impact on the
proposed algorithms of different initializations . This study
confirmed that the algorithms are not subject to performance
degradation due to the initialization although we have to bear in
mind that the initialization ${\boldsymbol T}_D(0)= {\boldsymbol
0}_{JM \times D}$ annihilates the signal and must be avoided. We
have also studied a particular case of the proposed scheme when
$JM=1$ and $D=1$, which yields the cost function
\begin{equation}
{\mathcal C}({\boldsymbol T}_D,\bar{\boldsymbol w} ) = E \big[ |d
- \bar{w} T_D r |^2 \big]  \big]
\end{equation}
By choosing $T_D$ (the "scalar" projection) fixed with $D$ equal
to $1$, it is evident that the resulting function ${\mathcal
C}(\bar{w},T_D=1,r) = |d-w^*~r|^{2} \big]$ is a convex one. In
contrast to that, for a time-varying projection $T_D$ the plots of
the function indicate that the function is no longer convex but it
also does not exhibit local minima. The problem at hand can be
generalized to the vector case, however, we can no longer verify
the existence of local minima due to the multi-dimensional
surface. This remains as an interesting open problem to be
studied.

\subsection{Proof of Convergence of the Method}

In this subsection, we show that the proposed reduced-rank
algorithm converges globally and exponentially to the optimal
reduced-rank estimator \cite{scharfo},\cite{hua1},\cite{hua2}. {
An issue that remains an open problem to be investigated is the
transient behavior of the proposed method, which corresponds to
the most significant difference between the proposed and existing
(MSWF and AVF) methods is on the transient performance. To our
knowledge, there exists no result for the transient analysis of
the MSWF and the AVF methods, even though it has been reported
(and also verified in our studies) that the AVF \cite{avf5} has a
superior convergence performance to the MSWF.}

As discussed in the previous subsection, the optimal solutions
${\boldsymbol T}_{D,{\rm opt}}$ and $\bar{\boldsymbol w}_{\rm
opt}$ are not unique. However, the desired product of the optimal
solutions, i.e., ${\boldsymbol w}_{\rm opt} = {\boldsymbol
T}_{D,{\rm opt}} \bar{\boldsymbol w}_{\rm opt}$ is known and given
by ${\boldsymbol R}^{-1/2} \big( {\boldsymbol
R}^{-1/2}{\boldsymbol p} \big)_{1:D}$
\cite{haykin},\cite{hua1},\cite{hua2}, where ${\boldsymbol
R}^{-1/2}$ is the square root of the input data covariance matrix
and the subscript ${1:D}$ denotes truncation of the subspace.

In order to proceed with our proof, let us rewrite the expressions
in (\ref{filtt}) and (\ref{filtw}) for time instant $0$ as follows
\begin{equation}
{\boldsymbol R}[0]{\boldsymbol T}_{D}[0] {\boldsymbol R}_{w}[0] =
{\boldsymbol P}_D[0] = {\boldsymbol p}[0]{\bar{\boldsymbol
w}}^H[0] , \label{filttalt}
\end{equation}
\begin{equation}
\bar{\boldsymbol R}[0]{\bar{\boldsymbol w}}[1] = {\boldsymbol
T}_{D}^H[0]{\boldsymbol R}[0]{\boldsymbol
T}_{D}[0]{\bar{\boldsymbol w}}[1] = \bar{\boldsymbol p}[0],
\label{filtwalt}
\end{equation}
Using (\ref{filttalt}) we can obtain the following relation
\begin{equation}
{\boldsymbol R}_{w}[0] = \big({\boldsymbol T}_{D}^H[0]{\boldsymbol
R}^2[0]{\boldsymbol T}_{D}[0]\big)^{-1} {\boldsymbol
T}_{D}^H[0]{\boldsymbol R}[0]{\boldsymbol p}[0]{\bar{\boldsymbol
w}}^H[0] , \label{filtrw}
\end{equation}
Substituting the above result for ${\boldsymbol R}_{w}[0]$ into
the expression in (\ref{filttalt}) we get a recursive expression
for ${\boldsymbol T}_{D}[0]$
\begin{equation}
\begin{split}
{\boldsymbol T}_{D}[0] & = {\boldsymbol R}[0]^{-1} {\boldsymbol
p}[0]{\bar{\boldsymbol w}}^H[0] \big( {\boldsymbol
T}_{D}^H[0]{\boldsymbol R}[0]{\boldsymbol p}[0]{\bar{\boldsymbol
w}}^H[0]\big)^{-1} \times \\ &  \quad  \times \big({\boldsymbol
T}_{D}^H[0]{\boldsymbol R}^2[0]{\boldsymbol T}_{D}[0]\big)^{-1},
\label{filtrwf}
\end{split}
\end{equation}
Using (\ref{filtwalt}) we can express $\bar{\boldsymbol w}[1]$ as
\begin{equation}
{\bar{\boldsymbol w}}[1] = \big({\boldsymbol
T}_{D}^H[0]{\boldsymbol R}[0]{\boldsymbol T}_{D}[0] \big)^{-1}
{\boldsymbol T}_{D}^H[0]{\boldsymbol p}[0], \label{filtwinst}
\end{equation}
Taking into account the relation ${\boldsymbol w}[1] =
{\boldsymbol T}_D[1] \bar{\boldsymbol w}[1]$, we obtain
\begin{equation}
\begin{split}
{{\boldsymbol w}}[1] & =  {\boldsymbol R}[1]^{-1} {\boldsymbol
p}[1]{\bar{\boldsymbol w}}^H[1] \big( {\boldsymbol
T}_{D}^H[1]{\boldsymbol R}[1]{\boldsymbol p}[1]{\bar{\boldsymbol
w}}^H[1]\big)^{-1} \cdot
\\ &  \quad  \cdot \big({\boldsymbol T}_{D}^H[1]{\boldsymbol
R}^2[1]{\boldsymbol T}_{D}[1]\big)^{-1}\big({\boldsymbol
T}_{D}^H[0]{\boldsymbol R}[0]{\boldsymbol T}_{D}[0] \big)^{-1}
{\boldsymbol T}_{D}^H[0]{\boldsymbol p}[0]
\end{split}
\end{equation}
More generally, we can express the proposed reduced-rank LS
algorithm by the following recursion
\begin{equation}
\begin{split}
{{\boldsymbol w}}[i] & = {\boldsymbol T}_D[i] \bar{\boldsymbol
w}[i] \\ & =  {\boldsymbol R}[i]^{-1} {\boldsymbol
p}[i]{\bar{\boldsymbol w}}^H[i] \big( {\boldsymbol
T}_{D}^H[i]{\boldsymbol R}[i]{\boldsymbol p}[i]{\bar{\boldsymbol
w}}^H[i]\big)^{-1} \cdot
\\ &  \quad  \cdot \big({\boldsymbol T}_{D}^H[i]{\boldsymbol
R}^2[i]{\boldsymbol T}_{D}[i]\big)^{-1} \cdot
\\ &  \quad  \cdot \big({\boldsymbol
T}_{D}^H[i-1]{\boldsymbol R}[i-1]{\boldsymbol T}_{D}[i-1]
\big)^{-1} {\boldsymbol T}_{D}^H[i-1]{\boldsymbol
p}[i-1]\label{filtwconc}.
\end{split}
\end{equation}
Since the optimal reduced-rank filter can be described by the SVD
of ${\boldsymbol R}^{-1/2}{\boldsymbol p}$ \cite{scharfo},
\cite{hua1},\cite{hua2}, where ${\boldsymbol R}^{-1/2}$ is the
square root of the covariance matrix ${\boldsymbol R}$ and
${\boldsymbol p}$ is the cross-correlation vector, then we have
\begin{equation}
{\boldsymbol R}^{-1/2}{\boldsymbol p} = {\boldsymbol \Phi}
{\boldsymbol \Lambda} {\boldsymbol \Phi}^H {\boldsymbol p}.
\label{optfilt}
\end{equation}
Considering that there exists some ${\boldsymbol w}[0]$ such that
the randomly selected ${\boldsymbol T}_D[0]$ can be written as
\cite{hua1},\cite{hua2}
\begin{equation}
{\boldsymbol T}_D[0] = {\boldsymbol R}^{-1/2}{\boldsymbol \Phi}
{\boldsymbol w}[0]. \label{tdo}
\end{equation}
Substituting (\ref{tdo}) and using (\ref{optfilt}) in
(\ref{filtwconc}), and manipulating the algebraic expressions, we
can express (\ref{filtwconc}) in a more compact way that is
suitable for analysis, as given by
\begin{equation}
{\boldsymbol w}[i] = {\boldsymbol \Lambda}^2 {\boldsymbol w}[i-1]
({\boldsymbol w}^H[i-1] {\boldsymbol \Lambda}^2  {\boldsymbol
w}[i-1] )^{-1} {\boldsymbol w}^H[i-1] {\boldsymbol w}[i-1].
\label{wdecomp}
\end{equation}
The above expression can be decomposed as follows
\begin{equation}
{\boldsymbol w}[i] = {\boldsymbol Q}[i]~{\boldsymbol
Q}[i-1]~\ldots~{\boldsymbol Q}[1]~{\boldsymbol w}[0], \label{wdec}
\end{equation}
where
\begin{equation}
{\boldsymbol Q}[i] = {\boldsymbol \Lambda}^{2i} {\boldsymbol
w}[0]( {\boldsymbol w}^H[0] {\boldsymbol \Lambda}^{4i-2}
{\boldsymbol w}[0])^{-1} {\boldsymbol w}^H[0] {\boldsymbol
\Lambda}^{2i-2} \label{qexp}.
\end{equation}
At this point, we need to establish that the norm of ${\boldsymbol
T}_D[i]$ for all $i$ is both lower and upper bounded, i.e., $0<
||{\boldsymbol T}_D[i]|| < \infty $ for all $i$, and that
${\boldsymbol w}[i] = {\boldsymbol T}_D[i] \bar{\boldsymbol w}[i]$
approaches ${\boldsymbol w}_{\rm opt}[i]$ exponentially as $i$
increases. Due to the linear mapping, the boundedness of
${\boldsymbol T}_D[i]$ is equivalent to that of ${\boldsymbol
w}[i]$. Therefore, we have upon convergence ${\boldsymbol
w}^H[i]{\boldsymbol w}[i-1] = {\boldsymbol w}^H[i-1] {\boldsymbol
w}[i-1]$. Since $||{\boldsymbol w}^H[i]{\boldsymbol w}[i-1]|| \leq
|| {\boldsymbol w}[i-1]|| || {\boldsymbol w}[i]||$ and
$||{\boldsymbol w}^H[i-1]{\boldsymbol w}[i-1]|| = || {\boldsymbol
w}[i-1]||^2$, the relation ${\boldsymbol w}^H[i]{\boldsymbol
w}[i-1] = {\boldsymbol w}^H[i-1] {\boldsymbol w}[i-1]$ implies
$||{\boldsymbol w}[i] || > ||{\boldsymbol w}[i-1]||$ and hence
\begin{equation}
||{\boldsymbol w}[\infty]|| \geq || {\boldsymbol w}[i]|| \geq
||{\boldsymbol w}[0]||
\end{equation}
In order to show that the upper bound $||{\boldsymbol
w}[\infty]||$ is finite, let us express the $JM \times JM$ matrix
${\boldsymbol Q}[i]$ as a function of the $JM \times 1$ vector
${\boldsymbol w}[i] = \left[
\begin{array}{c} {\boldsymbol w}_1[i] \\ {\boldsymbol w}_2[i] \end{array} \right]$
and the $JM \times JM$ matrix $ {\boldsymbol \Lambda} = \left[
\begin{array}{cc} {\boldsymbol
\Lambda}_1 & \\
& {\boldsymbol \Lambda}_2 \end{array} \right]$. Substituting the
previous expressions of ${\boldsymbol w}[i]$ and ${\boldsymbol
\Lambda}$ into ${\boldsymbol Q}[i]$ given in (\ref{qexp}), we
obtain
\begin{equation}
\begin{split}
{\boldsymbol Q}[i] & = \left[
\begin{array}{c} {\boldsymbol
\Lambda}_1^{2i} {\boldsymbol w}_1[0] \\ {\boldsymbol
\Lambda}_2^{2i} {\boldsymbol w}_2[0] \end{array} \right] (
{\boldsymbol w}^H_1[0] {\boldsymbol \Lambda}^{4i-2}_1 {\boldsymbol
w}_1[0] \\ & \quad + {\boldsymbol w}^H_2[0] {\boldsymbol
\Lambda}^{4i-2}_2 {\boldsymbol w}_2[0])^{-1} \left[
\begin{array}{c} {\boldsymbol w}_1^H[0]{\boldsymbol
\Lambda}_1^{2i-2}  \\  {\boldsymbol w}_2^H[0]{\boldsymbol
\Lambda}_2^{2i-2} \end{array} \right] \label{qexp2}.
\end{split}
\end{equation}
Applying the matrix identity $({\boldsymbol A} + {\boldsymbol
B})^{-1} = {\boldsymbol A}^{-1} - {\boldsymbol A}^{-1}{\boldsymbol
B} ({\boldsymbol I} + {\boldsymbol A}^{-1}{\boldsymbol B})^{-1}
{\boldsymbol A}^{-1}$ to the decomposed ${\boldsymbol Q}[i]$ in
(\ref{qexp2}) and making $i$ large, we get
\begin{equation}
{\boldsymbol Q}[i] = {\rm diag} \big( \underbrace{1 \ldots 1}_{D}~
\underbrace{0 \ldots 0}_{JM-D} \big) + {\rm O}(\epsilon[i]).
\label{Qres}
\end{equation}
where $\epsilon[i] = (\lambda_{r+1}/ \lambda_{r})^{2i}$ with
$\lambda_{r+1}$ and $\lambda_r$ are the $(r+1)$th and the $r$th
largest singular values of ${\boldsymbol R}^{-1/2}{\boldsymbol
p}$. From (\ref{Qres}), it follows that for some positive constant
$k$, we have $||{\boldsymbol w}[i]|| \leq 1 + k \epsilon[i]$. From
(\ref{wdec}), we obtain
\begin{equation}
\begin{split}
||{\boldsymbol w}[\infty]|| & \leq ||{\boldsymbol Q}[\infty]||
\ldots ||{\boldsymbol Q}[2]||~||{\boldsymbol Q}[1]||~ ||
{\boldsymbol Q}[0]|| \\ & \leq ||{\boldsymbol w}[0]||
\prod_{i=1}^{\infty} (1+ k \epsilon[i]) \\
& = ||{\boldsymbol w}[0]|| \exp \Big(\sum_{i=1}^{\infty} log (1+ k
\epsilon[i])\Big) \\
& \leq ||{\boldsymbol w}[0]|| \exp \Big( \sum_{i=1}^{\infty} k
\epsilon[i] \Big) \\ &  = ||{\boldsymbol w}[0]|| \exp \Big(
\frac{k}{1- (\lambda_{r+1}/\lambda_r)^2 } \Big)
\end{split}
\end{equation}
With the development above, the norm of ${\boldsymbol w}[i]$ is
proven to be both lower and upper bounded. Once this is
established, the expression in (\ref{filtwconc}) converges for
large $i$ to the reduced-rank Wiener filter. This can be verified
by equating the terms of (\ref{wdecomp}), which yields
\begin{equation}
\begin{split}
{{\boldsymbol w}}[i] & =  {\boldsymbol R}[i]^{-1} {\boldsymbol
p}[i]{\bar{\boldsymbol w}}^H[i] \big( {\boldsymbol
T}_{D}^H[i]{\boldsymbol R}[i]{\boldsymbol p}[i]{\bar{\boldsymbol
w}}^H[i]\big)^{-1}  \big({\boldsymbol T}_{D}^H[i]{\boldsymbol
R}^2[i]{\boldsymbol T}_{D}[i]\big)^{-1} \cdot
\\ &  \quad  \cdot \big({\boldsymbol
T}_{D}^H[i-1]{\boldsymbol R}[i-1]{\boldsymbol T}_{D}[i-1]
\big)^{-1} {\boldsymbol T}_{D}^H[i-1]{\boldsymbol p}[i-1] \\
&  = {\boldsymbol R}^{-1/2} {\boldsymbol \Phi}_1 {\boldsymbol
\Lambda}_1 {\boldsymbol \Phi}_1^{H}{\boldsymbol p} + {\rm
O}(\epsilon[i]) \label{filtwconc2}.
\end{split}
\end{equation}
where ${\boldsymbol \Phi}_1$ is a $JM \times D$ matrix with the
$D$ largest eigenvectors of ${\boldsymbol R}$ and ${\boldsymbol
\Lambda}_1$ is a $D \times D$ matrix with the largest eigenvalues
of ${\boldsymbol R}$.

\section{Simulations}

The performance of the proposed scheme and algorithms  is assessed
in terms of the uncoded bit error rate (BER) via simulations for
space-time interference suppression in a DS-CDMA system. {  We
consider dynamic fading situations, perfect synchronization and
the proposed and existing adaptive algorithms are employed to
adjust the filters and track the channel variations.}
Specifically, in our proposed reduced-rank STAP the output of the
receiver $x[i]$ is the input to a slicer that makes the decision
about the transmitted symbol $\hat{b}_k[i]$ for user $k$ as
follows
\begin{equation}
\hat{b}_k[i] = Q \big( x[i] \big) = Q \big( \hat{\bf w}^H[i]
{\boldsymbol T}_D^H[i] {\boldsymbol r}[i] \big)
\end{equation}
where $Q\big( \cdot \big)$ is the function that implements the
slicer and the $k$th user is assumed to be user $1$.

{For all simulations, we use the initial values $\bar{\boldsymbol
w}[0]= [1~0~\ldots~0]^T$ and ${\boldsymbol T}_D[0]=[{\boldsymbol
I}_D ~ {\boldsymbol 0}_{D,JM-D}]^T$. We assume $L=9$ as an upper
bound, employ QPSK symbols and $3$-path channels {  with a power
delay profile \cite{rappa} given by $0$, $-3$ and $-6$ dB}, where
in each run the spacing between paths is obtained from a discrete
uniform random variable between $1$ and $2$ chips and the
experiments are averaged over $200$ runs. The power and the phase
of each path is time-varying and follows Clarke's model
\cite{rappa}. This procedure corresponds to the generation of
independent sequences of correlated unit power Rayleigh random
variables for each path. The DoAs of the interferers and the
desired user are uniformly distributed in $(0,2\pi/3)$. The system
has a power distribution among the users for each run that follows
a log-normal distribution with associated standard deviation equal
to $1.5$ dB. We compare the proposed scheme with the Full-rank
\cite{haykin}, the MSWF \cite{goldstein} and the AVF \cite{avf}
techniques for the design of linear space-time receivers and also
the rank selection algorithms reported in \cite{goldstein} and the
\cite{avf5} with the proposed rank selection techniques.}

\begin{figure}[!htb]
\begin{center}
\def\epsfsize#1#2{1\columnwidth}
\epsfbox{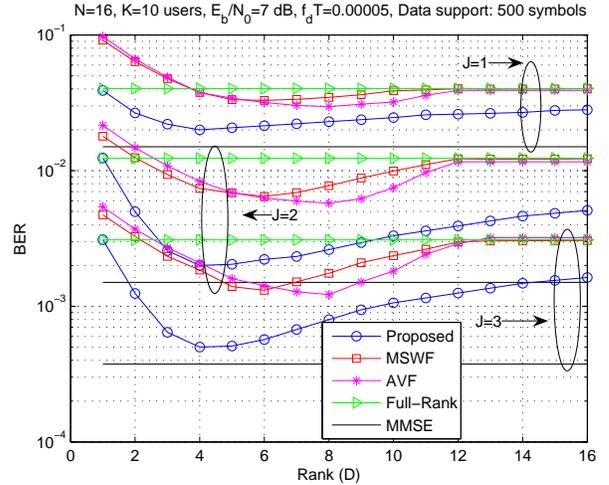} \caption{\small BER performance versus rank (D).}
\label{berxrank}
\end{center}
\end{figure}

In the first scenario, we consider the BER performance versus the
rank $D$ with optimized parameters (forgetting factors
$\lambda=0.998$) for all schemes. The results in Fig.
\ref{berxrank} indicate that the best rank for the proposed scheme
is $D=4$ for a data record of $500$ symbols as it is very close to
the optimal linear MMSE estimator. Studies with systems with
different processing gains and loads show that $D$ does not vary
significantly with either the system size or the load. However, it
should be remarked that considerable performance gains can be
obtained with an automatic rank adaptation algorithm for fine
tuning the used rank.

\begin{figure}[!htb]
\begin{center}
\def\epsfsize#1#2{1\columnwidth}
\epsfbox{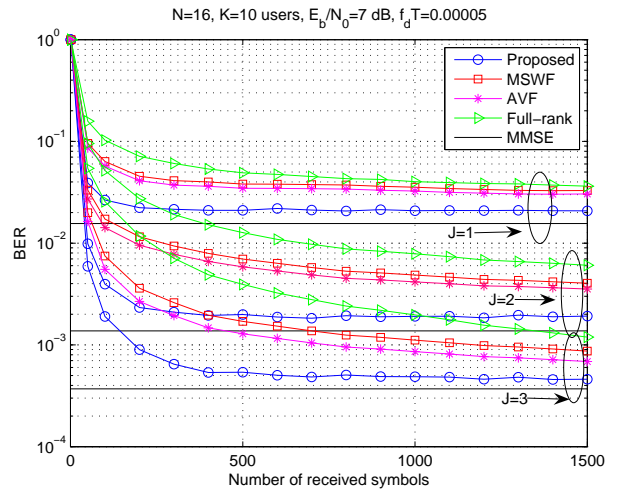} \caption{\small BER performance versus number of
received symbols.} \label{berxsym}
\end{center}
\end{figure}

In a second experiment, the BER convergence performance in a
mobile communications situation is shown in Fig. \ref{berxsym}.
The channel coefficients are obtained with Clarke´s model
\cite{rappa} and the adaptive estimators of all methods are
trained with $200$ symbols and are then switched to
decision-directed mode. The results show that the proposed scheme
has considerably better performance than the existing approaches
and is able to adequately track the desired signal. {  In
particular, the proposed reduced-rank algorithm converges in $100$
symbols for the case of $J=1$, in about $200$ symbols for the case
of $J=2$ and in about $400$ symbols for $J=3$.} This is
substantially faster than the existing reduced-rank schemes,
namely, the MSWF and the AVF (which are known to have the best
performances available in the area) and the full-rank RLS
algorithm.

\begin{figure}[!htb]
\begin{center}
\def\epsfsize#1#2{1\columnwidth}
\epsfbox{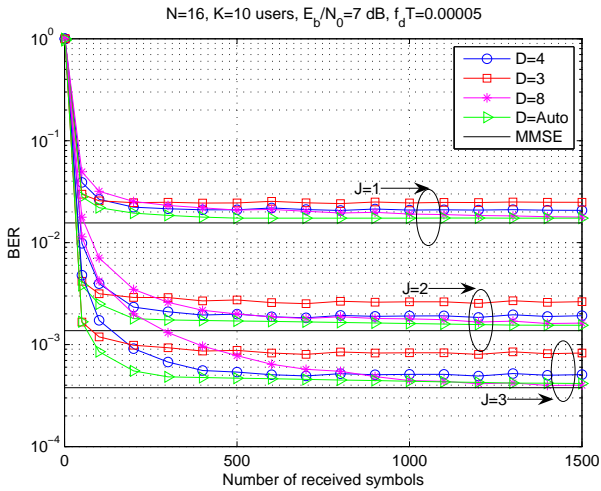} \caption{\small BER performance versus number of
received symbols with automatic rank adaptation.} \label{rankauto}
\end{center}
\end{figure}

In practice, the rank $D$ can be adapted in order to obtain fast
convergence and ensure good steady state performance and tracking
after convergence. To this end, we developed the automatic rank
selection algorithm in Section V. We will assess this algorithm in
a scenario identical to the previous experiment. The results in
Fig. \ref{rankauto} show that significant gains can be obtained
from the use of the automatic rank selection algorithm.
Specifically, we can notice that the proposed reduced-rank
algorithm has a very fast convergence with $D=3$ even though it
does not provide a steady state performance close to the full-rank
optimal linear MMSE estimator. {  When the proposed reduced-rank
algorithm employs $D=8$ the convergence is notably slower even
though it is able to approach the full-rank optimal linear MMSE
estimator in steady state as shown in Fig. \ref{rankauto} and
evidenced in our studies.} Interestingly, when equipped with the
proposed automatic rank selection algorithm the proposed
reduced-rank RLS algorithm achieves a convergence performance as
good as with $D=3$ and a steady state performance equivalent to
that with $D=8$. Another important issue is that the differences
in performance are more pronounced for larger filters, when the
usefulness of the automatic rank selection algorithm becomes more
clear.

\begin{figure}[!htb]
\begin{center}
\def\epsfsize#1#2{1\columnwidth}
\epsfbox{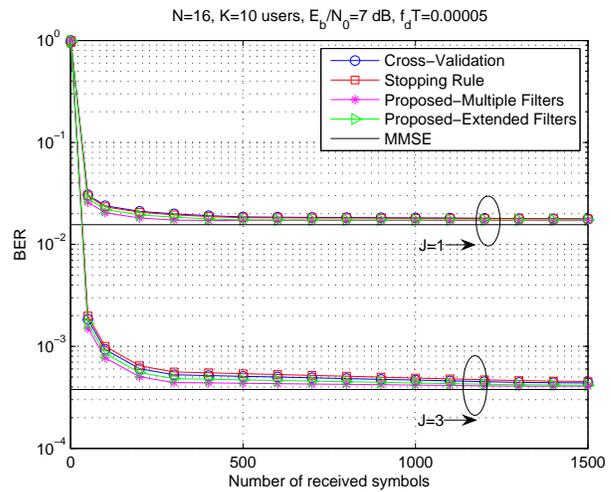} \caption{\small BER performance versus number of
received symbols with different automatic rank adaptation algorithms
and the proposed reduced-rank scheme and algorithm.}
\label{rankauto2}
\end{center}
\end{figure}

{In order to assess the performance of the proposed rank selection
algorithms, we consider the scenario of the previous experiment
with $J=1,3$ and compare the rank selection algorithms based on a
stopping criterion \cite{goldstein}, the cross-validation method
of \cite{avf5} and the proposed LS-based method with two
variations, namely, the multiple filters and the extended filters
approaches. The results shown in Fig. \ref{rankauto2} indicate
that the LS-based methods are slightly better than the other
techniques. The cross-validation approach has the advantage that
it does not require the setting of $D_{\rm min}$ and $D_{\rm
max}$, however, it may perform a search over a higher range of
values that leads to higher complexity. The remaining techniques
operate with $D_{\rm min}=3$ and $D_{\rm max}=8$. The method with
a stopping rule has a performance slightly worse than the
remaining schemes and its complexity is higher than the LS-based
techniques due to the computation of the orthogonal projection.  }

\begin{figure}[!htb]
\begin{center}
\def\epsfsize#1#2{1\columnwidth}
\epsfbox{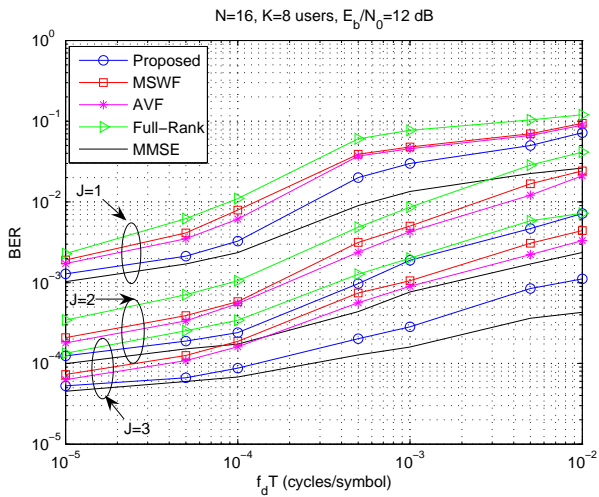} \caption{\small BER performance versus number of
received symbols. } \label{berxfdt}
\end{center}
\end{figure}

At this point, we will consider a study of the BER performance
against the normalized fading rate of the channel ($f_dT$) in the
experiment shown in Fig. \ref{berxfdt}. We assess the performance
of the receivers with a data record of $1000$ symbols of training.
{The proposed algorithm is equipped with the automatic rank
selection algorithm and the MSWF and the AVF algorithms are also
equipped with the rank adaptation techniques reported in
\cite{goldstein} and \cite{avf5}, respectively.} We observe from
the curves in Fig. \ref{berxfdt} that the proposed reduced-rank
algorithm obtains substantial gains in BER performance over the
existing MSWF and AVF algorithms and the full-rank RLS algorithm.
We can notice that as the channel becomes more hostile the
performance of the analyzed algorithms degrade, indicating that
the adaptive techniques are encountering difficulties in dealing
with the changing environment and interference. This behavior is
more pronounced when the algorithms have to adjust filters with
more coefficients, e.g. for more antenna elements ($J=2,3$). In
this regard, the reduced-rank algorithms obtain significant gains
over the full-rank RLS algorithm and, in particular, the proposed
reduced-rank algorithm achieves the best performance among them.

\begin{figure}[!htb]
\begin{center}
\def\epsfsize#1#2{1\columnwidth}
\epsfbox{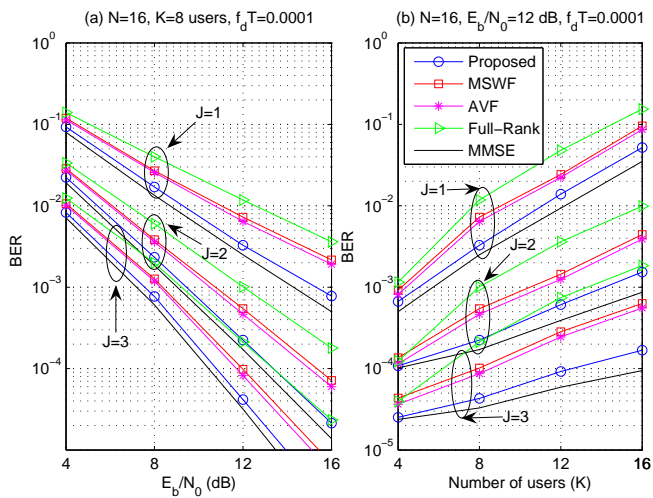} \caption{BER performance against (a) $E_b/N_0$
(dB) and (b) Number of Users (K) for different techniques.}
\label{berxsnr&k}
\end{center}
\end{figure}

The last experiment shows the BER performance versus the $E_b/N_0$
and the number of users ($K$), which is illustrated in Fig.
\ref{berxsnr&k}. In this scenario, all algorithms are trained with
$200$ symbols and are switched to decision-directed mode for
processing another $1500$ symbols. The curves are obtained after
$5000$ runs. The proposed algorithm is equipped with the automatic
rank selection algorithm and the MSWF and the AVF techniques are
also equipped with the rank adaptation methods reported in
\cite{goldstein} and \cite{avf5}, respectively. The results
confirm the excellent performance of the proposed reduced-rank
algorithm, which can approach the performance of the optimal MMSE
full-rank linear estimator (denoted simply as MMSE) that assumes
the knowledge of the channels, the DoAs and the noise variance. In
particular, the proposed reduced-rank algorithm can save up to $2$
dB in $E_b/N_0$ in comparison with the existing reduced-rank
techniques for the same BER performance, whereas it can
accommodate up to $4$ more users than the MSWF and the AVF for the
same BER performance. Interestingly, the performance of the
optimal reduced-rank linear MMSE estimator \cite{scharfo} that
assumes the knowledge of ${\boldsymbol R}$ and employs SVD is
quite similar to the optimal full-rank one. For this reason, we
only show the performance of the full-rank optimal linear MMSE
estimator.

\section{Conclusions}

We proposed a reduced-rank scheme based on joint iterative
optimization of parameter vectors. In the proposed scheme, the
full-rank adaptive filters are responsible for estimating the
subspace projection rather than the desired signal, which is
estimated by a small reduced-rank filter. We developed a
computationally efficient RLS algorithm for estimating the
parameters of the proposed scheme and an automatic rank selection
algorithm for computing the rank of the proposed RLS algorithm.
The proposed algorithms do not require an SVD for dimensionality
reduction and any knowledge about the order of the reduced-rank
model. The results for space-time interference suppression in a
DS-CDMA system show a performance significantly better than
existing schemes and close to the full-rank optimal linear MMSE
estimator in dynamic and hostile environments. The proposed
algorithms can be employed in a variety of applications including
spread spectrum and MIMO systems, wireless networks, cooperative
communications and navigation receivers.

\end{document}